\documentclass[aps,showpacs,prd,twocolumn]{revtex4}

\IfFileExists{srcltx.sty}{\usepackage[active]{srcltx}}

\usepackage{epsfig}
\usepackage{graphicx}
\usepackage{bm}
\usepackage{amsmath}
\usepackage{amssymb}

\newcommand{\beq}{\begin{equation}}
\newcommand{\eeq}{\end{equation}}
\newcommand{\bea}{\begin{eqnarray}}
\newcommand{\eea}{\end{eqnarray}}

\def\keV{\: {\rm keV}}

\def\GeV{\: {\rm GeV}}

\newcommand{\slashed}[1]{{#1}\hspace{-2mm}/}

\def\simle{\lower 2pt \hbox {$\buildrel < \over {\scriptstyle \sim }$}}
\def\simge{\lower 2pt \hbox {$\buildrel > \over {\scriptstyle \sim }$}}

\begin{document}

\title{Small-scale structure formation properties of chilled sterile neutrinos as dark matter.}

\author{Kalliopi Petraki}

\affiliation{Department of Physics and Astronomy, University of California, Los Angeles, California 90095-1547, USA}

%\date{}

\begin{abstract}
We calculate the free-streaming length and the phase space density of dark-matter sterile neutrinos produced from decays, at the electroweak scale, of a gauge singlet in the Higgs sector. These quantities, which depend on the dark-matter production mechanism, are relevant to the study of small-scale structure formation and may be used to constrain or rule out dark-matter candidates.

\end{abstract}

\pacs{14.60.St, 95.35.+d,   \hfill UCLA/08/TEP/02}

%\date{}

\maketitle

\section{Introduction}

The sterile neutrino is an appealing dark-matter candidate.  The gauge singlet fermions are universally used in models of neutrino masses\cite{review}; the seesaw Lagrangian~\cite{seesaw} built out of gauge singlet and non-singlet neutrinos can explain the observed neutrino masses and mixings for a wide range of Majorana masses.  If some of these masses are small, the corresponding new degrees of freedom appear in the low-energy effective theory as sterile neutrinos.  Sterile neutrinos with masses in the keV range can account for cosmological dark matter~\cite{dw,shi_fuller,dm_s,shaposhnikov_tkachev,Kusenko:2006rh, Petraki:2007gq} and can explain the observed velocities of pulsars~\cite{pulsars}.  They can also play a key role in baryogenesis~\cite{baryogenesis} and in the formation of the first stars~\cite{reion}.  Unlike many other candidates for dark matter, the sterile neutrinos have a nonzero free-streaming length that depends on their mass and the production history~\cite{Asaka:2006ek,Kusenko:2006rh,Petraki:2007gq}.  The non-negligible free streaming has an observable effect on the structure formation.  In this paper we will quantify the related properties of dark matter in the form of sterile neutrinos produced at the electroweak scale.

Dark matter with a negligibly small free-streaming length, called cold dark matter (CDM), is consistent with the observations of large-scale structure and microwave anisotropy.  Warm dark matter (WDM), with a free-streaming length smaller than 1~Mpc, fits the large-scale structure equally well.  The difference arises on the smaller scales.  CDM predictions for the structure  on small scales have been studied numerically, and a number of inconsistencies have been reported between the predictions of CDM and the observations~\cite{cdm-wdm,Boyanovsky:2007ay,Dalcanton:2000hn}.  In contrast, WDM solves these problems by suppressing the structure on the small scales.  It is possible that these discrepancies could go away as the simulations and the observations further improve. However, it is also possible that we are seeing the hints of dark matter in the form of sterile neutrinos.

In this paper, we focus on the small-scale structure formation properties of dark-matter sterile neutrinos, produced, at the electroweak scale, from decays of a gauge singlet in the Higgs sector. The Higgs singlet vacuum expectation value (VEV) gives rise to the Majorana masses of sterile neutrinos~\cite{Chikashige:1980ht}.
Decays of a gauge singlet Higgs, playing also the role of the inflaton, were proposed as the origin of the relic population of sterile neutrinos in Ref.~\cite{shaposhnikov_tkachev}. The decays occur around the mass scale of the Higgs singlet, which, in that model, is taken to be in the sub-GeV range. The VEV of the singlet Higgs varies greatly from its mass.
Here, however, we consider production of sterile neutrinos through decays of a singlet Higgs with mass and VEV both in the same energy scale~\cite{Kusenko:2006rh,Petraki:2007gq}. The requirement that a keV sterile neutrino constitutes all dark matter fixes this scale in the electroweak range, resulting in dark-matter particles with average momentum 3 times lower than in the Dodelson--Widrow production mechanism~\cite{dw}. In what follows, we will briefly review the definitions of free-streaming length and phase space density, and we will discuss their relevance to structure formation. We will then calculate these quantities for the suggested mechanism. The results can be compared to observations.

\section{Free-Streaming and Phase Space Density}

The free-streaming length $\lambda_{\rm fs}$ is the scale that corresponds to the cutoff of the power spectrum of the density perturbations. In addition to suppressing the structure on small scales, it may determine the filamentary structures in which the first stars form~\cite{Gao:2007yk}.  The free-streaming length is usually taken to be the comoving distance traveled by a particle within a Hubble time~\cite{kt}.  More detailed studies~\cite{kfs} have shown that $\lambda_{\rm fs}$ is the scale of the excitation mode that defines the crossover between those perturbations that grow under gravitational instability and those that are Landau damped. Boyanovsky~\cite{Boyanovsky:2007ba} calculated the marginal comoving  wavevector $k_{\rm fs}$, for multicomponent dark matter, consisting of several species of collisionless particles that interact only gravitationally. At  redshift $z$, $k_{\rm fs}(z)$ is
\beq
k_{\rm fs}^2(z) = \frac{4\pi G}{1+z} \sum_a \rho_a \left\langle\frac{1}{p^2/m^2}\right\rangle_a
\label{kfsdef}
\eeq
where the summation extends over all species constituting dark matter. In (\ref{kfsdef}), all of the dependence on the redshift $z$ is included in the prefactor $(1+z)^{-1}$. The mass density $\rho$ and $\langle p^{-2} \rangle$ are meant to be evaluated at the present time.

Here, we consider dark matter consisting of only one component. We define $x=p/T$ to be the comoving
momentum. Then, in terms of the distribution function of the dark-matter population $n(x)$,
immediately after the production has been completed, the free-streaming length, calculated at the present epoch ($z=0$), $\lambda_{\rm fs}~=~\frac{2\pi}{k_{\rm fs}(0)}$, is
\bea
\lambda_{\rm fs} &=&  8 \cdot 10^{-3} \; \frac{1}{\xi^{\frac{1}{3}}} \left(\frac{\int_0^\infty x^2 n(x) dx}{\int_0^\infty n(x) dx} \right)^{\frac{1}{2}} \times \nonumber \\
&& \left(\frac{0.2}{\Omega_d}\right)^{\frac{1}{2}} \left(\frac{\keV}{m}\right) \; {\rm Mpc}
\label{lfs}
\eea
where the factor $\xi=\frac{g_*(T_{\rm prod})}{g_*(T_{\rm today})}$ accounts for the dilution and the redshift of the dark-matter particles, caused by the entropy release due to the decoupling of relativistic degrees of freedom since the time of production at $T=T_{\rm prod}$.

On the observational side, the Lyman-$\alpha$ forest power spectrum as measured by the Sloan Digital Sky Survey (SDSS) yields a limit of $k_{\rm fs}~\geqslant~2~{\rm Mpc}^{-1}$ or $\lambda_{\rm fs} \leqslant 3 \; {\rm Mpc}$ at redshifts $z\approx 2-6$, which translates into $\lambda_{\rm fs} \leqslant 1.1 \; {\rm Mpc}$ today ($z=0$)~\cite{viel}.

\medskip

In order to study the phase-packing that leads to the appearance of central cores in galaxy halos, the phase space density, first introduced by Tremaine and Gunn~\cite{Tremaine:1979we}, is often employed. One has to discern between the fine-grained phase density, which is the distribution function, and the coarse-grained one. Boyanovsky \textit{et al.}~\cite{Boyanovsky:2007ay} generalized the coarse-grained phase density for arbitrary distribution functions of dark-matter particles. Here we will adopt, up to a change in units, this definition:
\beq
Q \equiv \frac{m N}{\left\langle\frac{p^2}{m^2}\right\rangle^{\frac{3}{2}}} = m^4 \mathcal{D}
\label{Qdef}
\eeq
where $N$ stands for the number density. $\mathcal{D} = N/\langle p^2\rangle^{3/2}$ is the dimensionless phase space density introduced in~\cite{Boyanovsky:2007ay}. For collisionless particles and in the absence of self-gravity, $Q$ (or $\mathcal{D}$) is a Liouville invariant quantity~\cite{Boyanovsky:2007ay}. Gravitational dynamics, however, can lead to decrease of the coarse-grained phase density. This signifies the entropy increase caused by dynamical heating during the non-linear gravitational clustering~\cite{Qdecrease-theory,Qdecrease-num,Dalcanton:2000hn,Hogan:2000bv}.
%, since $S \propto - \log Q$. (For a uniform monoatomic ideal thermal gas consisting of N particles: $S = -k_B N \log Q + const$)

The value of the primordial phase density $Q_0$ sets an upper limit to the dark-matter density $\rho$, for a given ``velocity'' dispersion, that is $\rho \leqslant 3^{3/2} Q_0 \sigma ^3$, where $\sigma~\equiv~\frac{1}{3^{1/2}}~\left\langle~\frac{p^2}{m^2}~\right\rangle^{\frac{1}{2}}$ is the 1-dimensional velocity dispersion in the nonrelativistic regime. For CDM, $Q_0\rightarrow \infty$, and there is no limit to the central density. For WDM, however, $Q_0$ is finite and halos form central cores rather than divergent cusps. Equivalently, this limit can be expressed in terms of the core radius $r_c$, which, in the case of an isothermal sphere model for the halo, is conventionally defined as $r_c = \sqrt{27 \sigma^2/4\pi G \rho}$. Then
\beq
r_c \geqslant \frac{\sqrt{3}}{4 \pi G} \frac{1}{\sqrt{Q_0 v_\infty}}
\label{r0} \eeq
where $v_\infty=\sqrt{6} \sigma$ is the asymptotic velocity of the halo's rotational curve.

Galaxy rotation curves provide a direct estimate of the enclosed mass density as a function of the distance from the galactic center, extending out to the flat portion. The latter yields an estimate for $v_\infty$, or equivalently the dark-matter velocity dispersion $\sigma$. Thus, rotation curves contain all the information needed to estimate the phase space density of the core. Recent observational data from the most dark matter dominated galaxies, the dwarf spheroidal galactic satellites, yield $Q_{\rm obs} \thickapprox 9 \cdot 10^{-6} - 2 \cdot 10^{-4} ~ \frac{M_{_{\bigodot}}/{\rm pc}^3}{\rm (km/s)^3}$~\cite{Q limits}.

One may expect that at least some portion of dark matter retains the low entropy state, corresponding to the primordial phase density, and is naturally concentrated in the center of the galaxy, forming its core. Simulations show that this is the case for CDM and it can still be a valid approximation for WDM, provided that cores are formed without a lot of dynamical heating. The relation between the core radius $r_c$ and the velocity dispersion $\sigma$, for a given dark-matter model, is a prediction of this assumption, which can be tested in halos. Independently of whether this approximation is good, the primordial phase density is bounded from below by the observable phase density, and this can be used to derive constraints for the mass of a dark-matter candidate.

The phase space density (\ref{Qdef}) can be expressed in terms of the dark matter distribution function $n(x)$:
\bea
Q &=& 1.8 \cdot 10^{-4} \; \xi \left(\frac{\int_0^\infty x^2 n(x) dx}{\int_0^\infty x^4 n(x) dx} \right)^{\frac{3}{2}} \times \nonumber \\
& & \left(\frac{\Omega_d}{0.2}\right) \left(\frac{m}{\keV}\right)^3 \; \frac{M_{_{\bigodot}}/{\rm pc}^3}{\rm (km/s)^3}
\label{Q}
\eea

\section{Sterile Neutrinos produced from singlet Higgs decays}

Sterile neutrinos can be produced through oscillations of the active neutrinos~\cite{dw}. This scenario appears to be in conflict with a combination of the X-ray bounds~\cite{x-rays} and the Lyman-$\alpha$ bounds~\cite{viel}. It is possible to evade this constraint, if the lepton asymmetry of the universe is greater than $O (10^{-3})$~\cite{shi_fuller}, or if decays of additional, heavier sterile neutrinos occur, introducing some additional entropy and contributing to cooling of dark matter~\cite{Asaka:2006ek}. More recent estimations of the free-streaming length and phase space density of sterile neutrinos produced via oscillations can be found in Refs.~\cite{Boyanovsky:2007ay,Boyanovsky:2007ba}.

An alternative production mechanism arises if the Higgs sector is extended by a (real) gauge singlet field $S$, which couples to the standard model Higgs, through a scalar potential, and to the right-handed neutrinos~\cite{shaposhnikov_tkachev,Kusenko:2006rh,Petraki:2007gq}
\bea {\cal L}  &=&  {\cal L_{\rm SM}} + i \bar N_a
\slashed{\partial} N_a - y_{\alpha a} H \,  \bar L_\alpha N_a -
\frac{f_a}{2} S \; \bar N_a^c N_a   \nonumber
\\ &-& V(H,S) + h.c. \label{LwS}
\eea
where
\bea V(H,S) &=& -\mu_H^2 |H|^2 - \frac{1}{2}\mu_S^2 S^2
+ \frac{1}{6}\alpha S^3 + \omega |H|^2 S  \nonumber
\\ &+& \lambda_H |H|^4 + \frac{1}{4}\lambda_S S^4 + 2\lambda_{HS}|H|^2 S^2
\label{V} \eea
In this model, the Majorana masses of right-handed neutrinos arise due to their Yukawa couplings to the $S$ boson, through spontaneous symmetry breaking. Sterile neutrinos can be produced from decays of the $S$ bosons. This possibility has been explored in detail in Refs~\cite{Kusenko:2006rh,Petraki:2007gq}. The requirement that the resulting abundance of a keV sterile neutrino constitutes all of the observed dark matter fixes the energy scale of $S$ bosons to be around the electroweak scale $\sim 10^2$~GeV. Sterile neutrinos are produced with a nonthermal spectrum and with average momentum lower than the equilibrium value at the same temperature. Since their production occurs at temperatures just a factor of a few below the $S$ boson mass, i.e. around the electroweak scale, sterile neutrinos are further redshifted by the entropy production due to the decoupling of the standard model degrees of freedom. We shall therefore call these sterile neutrinos ``Chilled Dark Matter". This chilling changes the quoted bounds for the sterile neutrino mass, derived from small-scale structure formation considerations~\cite{viel}. In what follows, we compute the potentially interesting, for this purpose, quantities of free-streaming length and phase space density, discussed in the previous section. As in Ref.~\cite{Petraki:2007gq}, we distinguish between two cases: in equilibrium and out-of-equilibrium singlet Higgs decay into sterile neutrinos.

\subsection{Decays in equilibrium}

If the couplings in the scalar potential (\ref{V}) of the Higgs singlet to the standard model Higgs are large enough, $S$ bosons remain in thermal equilibrium down to low temperatures, and decay while in equilibrium. The distribution function of dark-matter sterile neutrinos produced from decays of the singlet Higgs, in thermal equilibrium, is~\cite{shaposhnikov_tkachev}
\beq
n^{\Theta}(x) = \frac{f^2 M_0}{3 \pi m_S} x^2 \int_1^\infty \frac{(z-1)^{3/2} dz}{e^{xz}-1}
\label{n-eq}
\eeq
where $f$ is the Yukawa coupling of the dark-matter sterile neutrino to the $S$ boson (indices dropped for simplicity), $m_S$ is the $S$ boson mass and $M_0~=~\left(\frac{45 M_{PL}^2}{4 \pi^3 g_*}\right)^{\frac{1}{2}}~\sim~10^{18}~\GeV$ is the reduced Planck mass. Since the decay rate of the S bosons peaks at temperature $T_{\rm prod}=m_S/r_{\rm prod}$, with $r_{\rm prod}=2.3$~\cite{Petraki:2007gq}, the sterile neutrino population will be diluted by $\xi \simeq 33$~\cite{Kusenko:2006rh}, representing the decoupling of all of the standard model degrees of freedom. Then, the free-streaming length and phase space density, Eqs. (\ref{lfs}) and (\ref{Q}), are
\bea
\lambda_{\rm fs}^{\Theta} &=&  2 \cdot 10^{-3} \left(\frac{33}{\xi}\right)^{\frac{1}{3}}  \left(\frac{0.2}{\Omega}\right)^{\frac{1}{2}} \left(\frac{\keV}{m}\right) \; {\rm Mpc}
\label{lfs-eq}
\\
Q^{\Theta} &=& 2.4 \cdot 10^{-4} \left(\frac{\xi}{33}\right) \left(\frac{\Omega}{0.2}\right) \left(\frac{m}{\keV}\right)^3 \; \frac{M_{_{\bigodot}}/{\rm pc}^3}{\rm (km/s)^3}
\label{D-eq}
\eea

\subsection{Out-of-equilibrium decays}
If $S$ bosons are weakly coupled to the standard model particles ($\alpha,\omega \approx 0, \; \lambda_{HS} \approx 10^{-6}$), they decouple early and decay at a later time, while out-of-equilibrium. In this case, the distribution function of the dark-matter sterile neutrinos produced is~\cite{Petraki:2007gq}:
\bea \lefteqn{n^{^{\displaystyle{\not} \Theta}}(x) = \frac{B}{x^2}
\left[ \int_{\left| \frac{r_{\rm f}^2}{4x}-x \right|}^\infty x_S n_S(x_S,r_{\rm
f}) dx_S  \right.} \nonumber
    \\ &-&  \left. \int_{r_{\rm f}}^\infty \frac{r}{2x}\left(\frac{r^2}{4x}-x\right)
n_S\left(\left|\frac{r^2}{4x}-x\right|,r\right)dr    \right]
\label{n-ooe-sol}   \eea
where $r=\frac{m_S}{T}$ and $n_S(x_S,r)$ is the $S$ boson distribution function after freeze-out occurs, at $r_{\rm f} = m_S/T_{\rm f}$:
\bea n_S(x_S,r) &=& \frac{1}{e^{\sqrt{x_S^2+r_{\rm f}^2}}-1}
\left(\frac{r+\sqrt{x_S^2+r^2}}{r_{\rm f}+\sqrt{x_S^2+r_{\rm
f}^2}}\right)^{\Lambda x_S^2} \times \nonumber \\
&& e^{-\Lambda (r \sqrt{x_S^2+r^2} - r_{\rm f} \sqrt{x_S^2+r_{\rm f}^2})}
\label{nS} \eea
\begin{figure}[b]
  \centering
  \includegraphics[width=6.5cm]{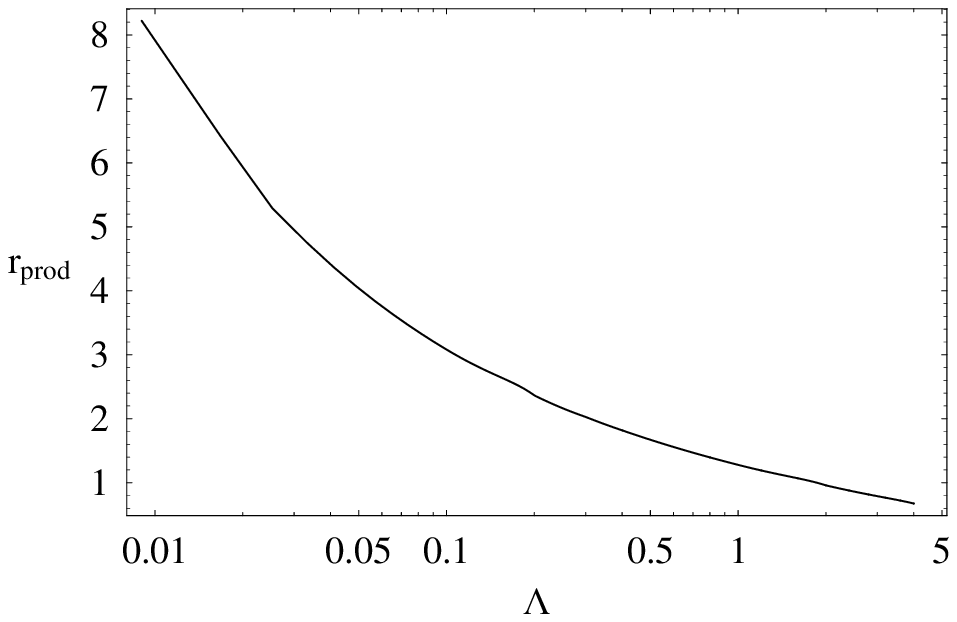}
  \caption{The dimensionless time parameter $r=\frac{m_S}{T}$, corresponding to the peak of the sterile neutrino production rate from out-of-equilibrium $S$ boson decays, vs $\Lambda$. For the range of $\Lambda$ considered, $S$ boson decays peak at temperatures only a factor of a few below its mass, that is before the decoupling of the QCD degrees of freedom.}
  \label{rprod}
\end{figure}
\begin{figure}[t]
  \centering
  \includegraphics[width=6.5cm]{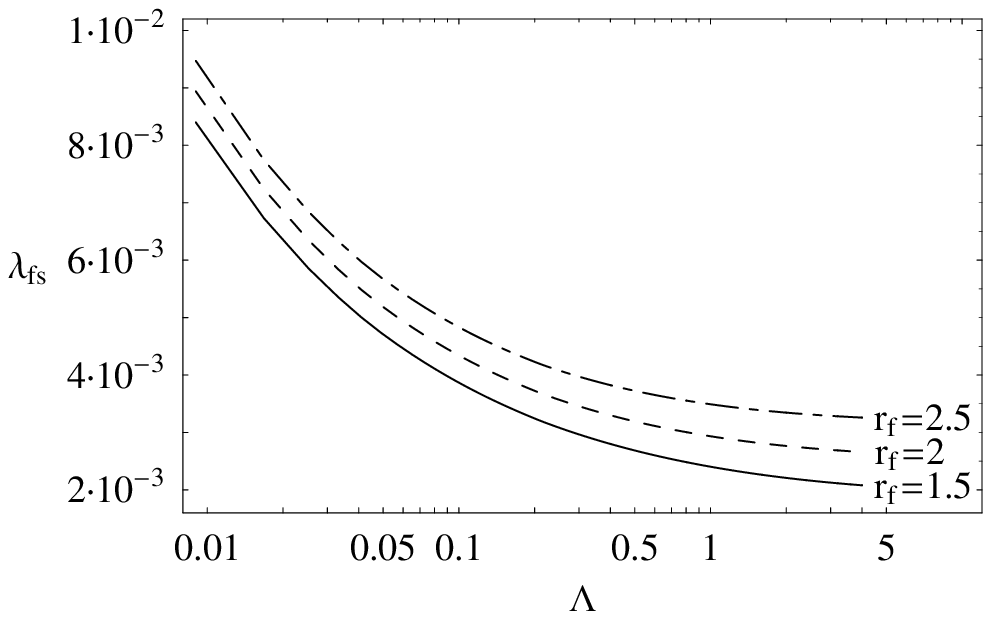}
  \includegraphics[width=6.5cm]{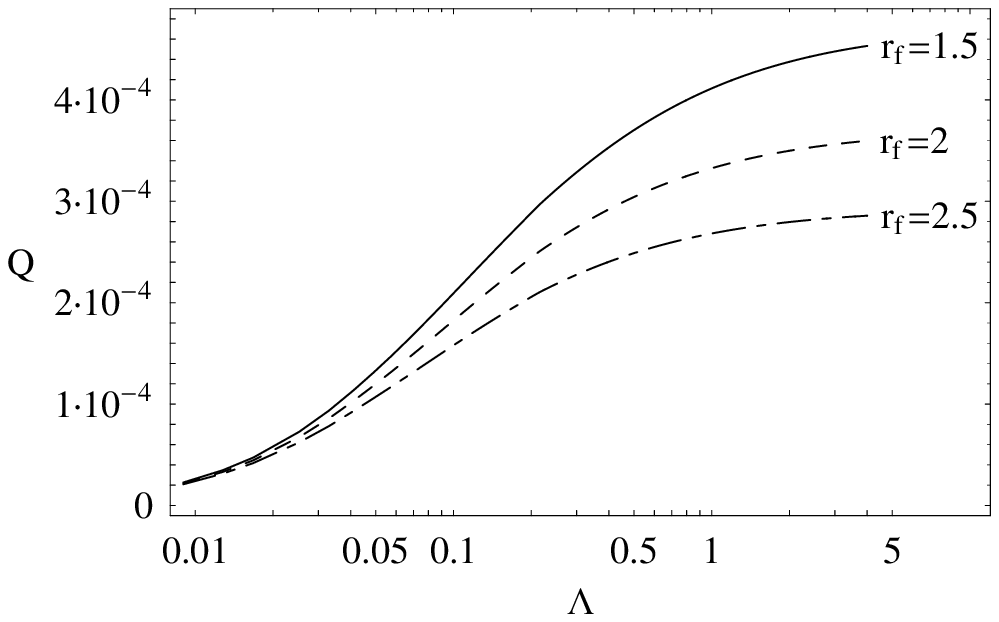}
  \caption{The free-streaming length and the phase space density for sterile neutrinos produced from out-of-equilibrium $S$ boson decays. $\lambda^{^{\displaystyle{\not} \Theta}}_{\rm fs}$ is in units of $\left(\frac{0.2}{\Omega}\right)^{\frac{1}{2}} \left(\frac{\keV}{m}\right) \; {\rm Mpc}$ and $Q^{^{\displaystyle{\not} \Theta}}$ is in units of $\left(\frac{\Omega}{0.2}\right) \left(\frac{m_s}{\keV}\right)^3 \; \frac{M_{_{\bigodot}}/{\rm pc}^3}{\rm (km/s)^3}$. Small $\Lambda$ implies delayed decays, resulting in warmer dark matter. Sterile neutrinos produced through this mechanism constitute all of dark matter for $\Lambda \approx 0.1$.}
  \label{Lfs,Q}
\end{figure}
Here, $\Lambda = \frac{h^2 M_0}{16 \pi m_S}$, where $h^2$ takes into account the decays of $S$ particles into all of the sterile neutrino species and the standard model fermions through the mass mixing with the standard model Higgs:
\bea
h^2 &\equiv& \sum_a f_a^2 \left(1-\frac{4f_a^2 \sigma^2}{m_S^2}\right) \nonumber
\\ &+& \sum_{\rm f} \lambda_f^2 \left(1-\frac{4m_f^2}{m_S^2}\right)
\left(\frac{\lambda_{HS}}{\max (\lambda_H,\lambda_S)}\right)^2
\label{h} \eea
$\sigma \sim 10^2$~GeV stands for the VEV of the $S$ boson and $B \equiv \frac{f^2}{h^2}$ is the branching ratio of $S$ decays into the dark-matter sterile neutrino.

The case of interest here is $\lambda_{HS} \approx 10^{-6}$. For lower values of $\lambda_{HS}$, $S$ bosons never come into equilibrium. This sets a lower limit on $\Lambda$, corresponding to the $b \bar{b}$ decay mode: $\Lambda \geqslant 0.01$ (where  we used $\lambda_H,\lambda_S < 1$, imposed by the requirement of perturbativity of the potential).  $\Lambda$ determines how fast $S$ particles decay. Small values of $\Lambda$ lead to delayed decays and therefore warmer dark matter. Here, however, the minimum value of $\Lambda$ ensures that $S$ decays peak at temperatures just a factor of a few below its mass, as shown in Fig.~\ref{rprod}. Since $S$ bosons live in the electroweak scale, the dark matter will be produced before the decoupling of the QCD degrees of freedom, which implies $\xi \thickapprox 25 - 33$. The free-streaming length and the phase space density of dark matter are shown in Fig.~\ref{Lfs,Q}. For definiteness, we used $m_S=200$ GeV. Out-of-equilibrium $S$ decays into a keV sterile neutrino produce the right amount of dark matter for $\Lambda \approx 0.1$~\cite{Petraki:2007gq}.

\section{Conclusions}

We have calculated the free-streaming length and the phase space density of dark matter in the form of sterile neutrinos produced from decays of a gauge singlet in the extended Higgs sector. These two quantities may not capture the entire spectrum of density perturbations, but they serve as a useful set of parameters for comparing the model predictions with the data.
We have considered two cases: in-equilibrium and out-of-equilibrium decays of the singlet Higgs. Comparison of the results, Eqs.~(\ref{lfs-eq}) and (\ref{D-eq}) with Fig.~\ref{Lfs,Q}, shows that in-equilibrium decays result in colder dark matter than out-of-equilibrium decays.

The quantities considered here have been calculated for other sterile neutrino production mechanisms in Refs.~\cite{Boyanovsky:2007ay,Boyanovsky:2007ba}. Sterile neutrinos produced from the singlet Higgs decays have lower free-streaming length and higher primordial phase space density than neutrinos produced nonresonantly via the Dodelson-Widrow mechanism~\cite{dw}. This is due to both their primordial nonthermal distribution and the chilling that occurs due to the entropy generation since their time of production. Compared to ``cool" sterile neutrinos produced by net-lepton number driven resonant conversion~\cite{shi_fuller}, chilled sterile neutrinos have comparable free-streaming length and lower phase density. Chilled sterile neutrinos can account for all of dark matter, which would be in agreement with the current observational limits. The Higgs structure of this model can be probed at the Large Hadron Collider at CERN and at a linear collider~\cite{singlet_higgs_LHC}.

\medskip

The author thanks D.~Boyanovsky and A.~Kusenko for helpful discussions.  This work was supported in
part by the DOE Grant No. DE-FG03-91ER40662 and by the NASA ATP Grant No. NAG~5-13399.

%\vspace{.2cm}
%Integrals used:
%\bea \int_0^\infty n(x) dx &=& 4B \int_0^{\infty} dx x^2  \left[\frac{1}{r_{\rm f}^2} n_S(x,r_{\rm f}) -2 \int_{r_{\rm %f}}^{\infty} \frac{dr}{r^3} n_S(x,r)  \right] \label{n int} \\
%\int_0^\infty x^2 n(x) dx &=& B \int_0^{\infty} dx x^2 n_S(x,r_{\rm f})  \label{x2n int} \\
%\int_0^\infty x^4 n(x) dx &=& \frac{B}{12} \int_0^{\infty} dx x^2  \left[(4 x^2 +3 r_{\rm f}^2) n_S(x,r_{\rm f})  +6 %\int_{r_{\rm f}}^{\infty} dr r n_S(x,r)  \right] \label{n int}
%\eea

\end{document}